\documentclass[12pt,a4paper]{article}
\usepackage[utf8x]{inputenc}
\usepackage[T1]{fontenc}
\usepackage{amsfonts}
\usepackage{amsmath}
\usepackage[pdftex]{graphicx} 
\usepackage[pdftex,linkcolor=black,pdfborder={0 0 0}]{hyperref} 
\usepackage{calc} 
\usepackage{enumitem} 
\usepackage{amsthm}
\usepackage{bm}
\usepackage{xcolor}
\usepackage[super]{natbib}
\usepackage{placeins}
\usepackage{comment}
\usepackage{algorithm2e}
\RestyleAlgo{ruled}
\usepackage{dsfont}
\usepackage{pdflscape}
\usepackage{pdfpages}

\def\spacingset#1{\renewcommand{\baselinestretch}%
{#1}\small\normalsize} \spacingset{1}

\DeclareMathOperator{\ND}{\mathcal{N}}
\DeclareMathOperator{\UD}{\mathcal{U}}

\newcommand\KL[2]{\mathcal{D}_\text{KL}\left\lbrack #1 \,\vert\vert\, #2\right\rbrack}

\def\diag{{\mbox{diag}}}

\def\wkl{w_k^{(l)}}
\def\Kl{K^{(l)}}
\def\zil{z_i^{(l)}}
\def\zilm1{z_i^{(l-1)}}
\def\mukl{\mu_k^{(l)}}
\def\Bkl{B_k^{(l)}}

\def\epsikl{\epsilon_{ik}^{(l)}}
\def\delkl{\Delta_k^{(l)}}
\def\Dl{D^{(l)}}
\def\Dlm1{D^{(l-1)}}
\def\delkl{\delta_k^{(l)}}

\def\thetaDMFA{\theta_{\text{DMFA}}}
\def\thetaReg{\theta_{\text{Reg}}}
\def\lambdaDMFA{{\lambda_{\text{DMFA}}}}
\def\lambdaReg{{\lambda_{\text{Reg}}}}

\DeclareMathOperator{\HCD}{\mathcal{HC}}
\DeclareMathOperator{\IGD}{\mathcal{IG}}

\begin{document} 

\begin{titlepage}

\title{Deep mixture of linear mixed models for complex longitudinal data}

\author{Lucas Kock$\mbox{}^1$, Nadja Klein$\mbox{}^2$ and David J. Nott$\mbox{}^1$} 

\date{\today}
\maketitle
\thispagestyle{empty}
\noindent
\vspace{2em}

\begin{center}
{\Large Abstract}
\end{center}
\vspace{-1pt}
\noindent Mixtures of linear mixed models are widely used for modelling longitudinal data for which observation
times differ between subjects. In typical applications, temporal trends are described using a basis expansion,
with basis coefficients treated as random effects varying by subject. Additional random effects can describe
variation between mixture components, or other known sources of variation in complex designs. A key
advantage of these models is that they provide a natural mechanism for clustering. Current versions of
mixtures of linear mixed models are not specifically designed for the case where there are many observations
per subject and complex temporal trends, which require a large number of basis functions to capture. In
this case, the subject-specific basis coefficients are a high-dimensional random effects vector, for which the
covariance matrix is hard to specify and estimate, especially if it varies between mixture components. To
address this issue, we consider the use of deep mixture of factor analyzers models as a prior for the random
effects. The resulting deep mixture of linear mixed models is well-suited for high-dimensional settings, and
we describe an efficient variational inference approach to posterior computation. The efficacy of the method
is demonstrated in biomedical applications and on simulated data.

\vspace{20pt}
 

\vspace*{\fill}
\noindent {\small\textbf{Acknowledgments:} Nadja Klein acknowledges support through the Emmy Noether grant KL 3037/1-1 of the German research foundation (DFG). The work of Lucas Kock and Nadja Klein was supported by the Volkswagenstiftung (grant: 96932). David Nott's research was supported by the Ministry of Education, Singapore, under the Academic Research Fund Tier 2 (MOE-T2EP20123-0009), and he is affiliated with the Institute of Operations Research and Analytics at the National University of Singapore.
}

\vspace{20pt}

\noindent{\small
$^1$ Department of Statistics and Data Science, National University of Singapore, Singapore\\
$^2$ Scientific Computing Center, Karlsruhe Institute of
Technology, Germany\\
$^\ast$ Correspondence should be directed to nadja.klein@kit.edu
}

\end{titlepage}

\spacingset{1.5}

\section{Introduction}

Longitudinal data play an important role in many biomedical applications \citep{RenTapFanChuTho2022,CaoAllVanGut2022}. In their practical use, mixtures of linear mixed models (MLMMs) \citep{verbeke+l96} are widely used for the analysis of longitudinal data for which observation times differ by subject, and in cases where there is a need to ``borrow strength" between subjects in a flexible way. A common approach to modelling temporal trends in MLMMs is to use flexible basis expansions, with basis coefficients treated as a random effect varying across individuals. The mixture structure for the distribution of the random effects provides flexibility when the random effects are  non-Gaussian, and also provides a natural mechanism for clustering which enhances interpretability. In settings where there are a large number of observations per subject and the temporal trends are complex, many basis functions may be required, which results in high-dimensional random effects.  The main contribution of this paper is to address the issue of high-dimensionality in MLMMs by using a  deep mixture of factor analyzers (DMFA) model as the prior for the random effects distribution. The result is a new deep mixture of linear mixed model (DMLMM) specification. We discuss efficient variational methods for computation and demonstrate the good performance of our approach in simulations and a number of real biomedical examples involving within subject prediction for unbalanced longitudinal biomarker data, likelihood-free inference (LFI) for modelling the temporal dynamics of malaria transmission and missing data imputation for gene expression data.  

A common application of MLMMs has been in clustering of time course gene expression data.  Several authors have considered linear mixed models (LMMs) with basis expansions for modelling temporal trends, and extensions to mixtures for clustering \citep{bar-joseph+ggjs02,luan+l03,qin+s06}.  A similar approach in the functional data analysis literature is described by James and Sugar\citep{james+s03}.  Celeux et al.\citep{celeux+ml05} consider MLMMs for clustering of gene expression datasets with replication, where gene level random effects are shared between replicates.  Ng et al.\citep{ng+mwjn06} extend this model with a random effect for different tissues, and Tan and Nott\citep{tan+n14} consider a similar model with two random effects,  one for subjects and one for the mixture component, and allow for covariate-dependent mixing weights. They consider Bayesian inference in their model, with computations carried out using variational approximation methods. Scharl et al.\citep{scharl+gl10} consider initialization of EM algorithms for mixtures of regression models, including MLMMs, for clustering  time series gene expression data. Pfeifer\citep{pfeifer04} clusters longitudinal data using LMMs, where the random effects distribution is either a finite mixture of normals, or some arbitrary distribution approximated discretely. Coke and Tsao\citep{coke+t10} consider clustering of electrical load series. MLMMs also arise in the literature on model-based functional clustering, where approximations to continuous time processes can lead to processes defined from finite basis expansions and a LMM formulation.  Examples include Chiou and Li\citep{chiou_functional_2007-1}, who consider a nonparametric random effects model and a truncated Karhunen-Lo\`{e}ve expansion, and Jacques and Preda\citep{jacques+p14} in which the authors cluster multivariate functional data assuming that multivariate functional principal components are normally distributed.   McDowell et al.\citep{mcdowell+mvsre18} perform functional clustering of gene expression data using a Dirichlet process Gaussian process  mixture model. Shi and Wang\citep{shi+w08} develop a mixture of Gaussian process functional regressions model where the mixing weights can be covariate-dependent.

There are a variety of generalizations or closely related models to finite MLMMs.  These include partition models \citep{heard+hs06,booth+ch08} mixtures of generalized LMMs (GLMMs)\citep{lenk+d00,Proust2005} and mixtures of nonlinear hierarchical models \citep{pauler+l00,delacruz-mesia+qm08}. Bai et al.\citep{Bai2016} robustify mixtures of linear mixed models by  assuming a multivariate-$t$ distribution for the responses and random effects jointly within each mixture component. LMMs with nonparametric priors, include infinite mixtures of LMMs or more general hierarchical models have been considered in the literature on Bayesian nonparametrics \citep{bush+m96,kleinman+i98,muller+r97,heinzl+t13}.

There have been several recent works integrating mixed effects models with deep learning methods. Kilian et al.\citep{KilYeKel2023} introduce techniques to introduce random effects post-hoc into arbitrary supervised regression models.  Tran et al.\citep{TraNguNotKoh2020} represent fixed and random effects of GLMMs through deep networks and use variational methods for inference in the resulting complex model. Similarly, Mandel et al.\citep{ManGhoBar2021} replace the linear effects of a mixed effects model with neural networks. The resulting model is especially suited to handle densely sampled longitudinal data. A recent overview on machine learning techniques for longitudinal biomedical data can be found in  Cascarano et al.\citep{CasMurHerCamDetGkoChaVitLek2023}.

Complementary to this existing literature, in our model, the DFMA introduced by Viroli and McLachlan\citep{viroli2019deep} serves as a prior for the random
effects in MLMMs.  It is based on a mixture of factor analyzers
model \citep{ghahramani+b00,mclachlan2003modelling} but instead
of assuming factors to be Gaussian, allows the factors to themselves
be modelled as a mixture of factor analyzers recursively
for multiple layers.  
The model of Viroli and McLachlan\citep{viroli2019deep} builds
on an earlier formulation described in Tang et al.\citep{Tang+sh2012}, 
where components are split recursively and the fitting is done
layerwise.  However, Viroli and McLachlan\citep{viroli2019deep} use 
a similar architecture to that in
van den Oord and Schrauwen\citep{Ord+s2014}, where the authors allow 
parameter sharing between
mixture components, although they do
not consider factor structures for the mixture component covariance
matrices.   Other related mixture models
are considered in Yang et al.\cite{yang+hz17}, Li\citep{Li2005} and 
 Malsiner-Walli et al.\citep{Malsiner-Walli+fg2017}.
We build on the Bayesian formulation of DMFAs proposed by Kock et al.\citep{KocKleNot2022} and implement efficient variational methods for computation.

The DMFA prior allows for complex high-dimensional random effects distributions. Conditional distributions derived from our DMLMM approach are analytically tractable, thus predictive distributions for unobserved time points can be derived in a computationally attractive manner. One scenario where this is useful is predictive LFI. Simulators with intractable likelihoods are commonly used in biomedical applications \citep{Tan2019,CleCouFelTra2015} and inference is often based on a large sample from the simulator. If each sample is a high-dimensional time series, a large number of basis functions is needed to estimate the temporal trend. Mixture models are a well established tool in LFI, where the goal is parameter inference \citep{BonYouWes2011,BonWes2015,ForNguNguArb2021}, but predictive LFI has not been explored within the MLMM literature before.

Throughout this paper, we demonstrate the adaptability of our DMLMM approach across a range of biomedical applications, each presenting distinct challenges in modern biostatistics. Firstly, we consider within subject prediction for an unbalanced longitudinal study. Secondly, we consider the task of predicting the number of malaria cases in Afghanistan based on an intractable simulator. Lastly, an application to missing data imputation for gene expression data is given within the online supplement. Mixture modelling allows adaptive local sharing of information which improves imputation. Across all these applications, the Gaussian mixture model (GMM) representation of the DMLMM  is helpful for interpretation and the derivation of additional insights. Python code for the DMLMM is publicly available \href{https://github.com/kocklucx/DMLMM}{github.com/kocklucx/DMLMM}.

The structure of the paper is as follows.  In the next section
we introduce the DMLMM for longitudinal data based on a Bayesian version of the DMFA model considered in Viroli and McLachlan\citep{viroli2019deep} and outline efficient variational inference methods for posterior estimation in Section \ref{sec:post}. Sections \ref{sec:app} and \ref{sec:sim} explore the properties of our approach in the aforementioned real applications and simulations. Section \ref{sec:disc} gives some concluding discussion.

\section{Deep mixture of LMMs}\label{sec:DMLMM}

This section introduces the DMLMM. Section~\ref{sec: model specification} describes the overall model, while Section~\ref{sec: dmfa} discusses the DMFA prior for the random effects in more detail.

\subsection{The DMLMM -- notation and model specification} \label{sec: model specification}

Consider a longitudinal study where data $y_i=(y_{i1},\dots, y_{in_i})^\top$ is observed for subject $i$, $i=1,\dots, n$, 
with $y_{ij}$ an observation at time $t_{ij}$, $j=1,\dots,n_i$. Writing $t_i=\left(t_{i1},\dots,t_{in_i}\right)^\top$, it is assumed that
\begin{align}\label{eq: regression layer}
    y_i=B(t_i)\beta_i+\varepsilon_i,
\end{align}
where $\varepsilon_i\sim\ND(0,\sigma^2I_{n_i})$, $B(t_i)=\left(B(t_{i1}),\dots,B(t_{in_i})\right)^\top$ is a known $n_i\times d$ design matrix where $B(t_{ij})$ is a $d$-dimensional column vector of basis functions evaluated at $t_{ij}$, and $\beta_i\in\mathbb{R}^d$ is a subject specific random coefficient vector. Note, that we do not enforce an explicit relationship between $n_i$ and $d$. In particular, we explicitly allow $n_i<d$ for some individuals $i$.  
We consider Bayesian inference, and use a half-Cauchy prior $\sigma\sim\HCD(A)$ for the standard deviation of the error terms $\varepsilon_i$, which we express hierarchically as 
\begin{align*}
& \sigma^2|\psi \sim \IGD\left(\frac{1}{2},\frac{1}{\psi}\right)\;\;\; \psi\sim \IGD\left(\frac{1}{2},\frac{1}{A^2}\right).
\end{align*}
We choose this thick-tailed prior for the error variance as it robustifies the model against conflicts with the data for example through outliers. Section \ref{sec: dmfa} introduces a DMFA model which 
we use as a flexible prior distribution for 
the random effects $\beta_i$.  Write $\beta=(\beta_1^\top,\dots, \beta_n^\top)^\top$, and $\theta=(\eta^\top,\beta^\top)$ where
$\theta$ are the unknown parameters, so that $\eta$ contains
the unknowns except for $\beta$.
The DMFA prior
density for $\beta_i$ is a GMM with density of the form
$$p(\beta_i|\eta)=\sum_{k=1}^K w_k \phi(\beta_i;\mu_k,\Sigma_k),$$ 
where $\sum w_k=1$, and $\phi(\cdot;\mu,\Sigma)$ denotes the multivariate normal density function with mean $\mu$ and covariance matrix $\Sigma$. In the DMFA the parameters $w_k$, $\mu_k$ and $\Sigma_k$ are parametrized parsimoniously and this is described
in detail later.  
Integrating out $\beta$ in \eqref{eq: regression layer}
using $p(\beta_i|\eta)$ gives the marginal likelihood 
\begin{align}\label{eq: gmm y}
   p(y_i|\eta) & = \sum_{k=1}^K w_k\phi(y_i;B(t_i)\mu_k,B(t_i)\Sigma_kB(t_i)^\top+\sigma^2I_{n_i}).
\end{align}
The random effects $\beta_i$ can be interpreted as projections of the unequal length observations $y_i$ into a joint $d$-dimensional latent space.  
Our later applications demonstrate that the flexible DMFA prior allows
complex trends to be modelled well when the number of
basis functions is large, while borrowing strength between
similar observations to stabilize estimation for subjects having little
available data.  

\begin{figure}[!ht]
\centering
\includegraphics[width=0.95\columnwidth,keepaspectratio]{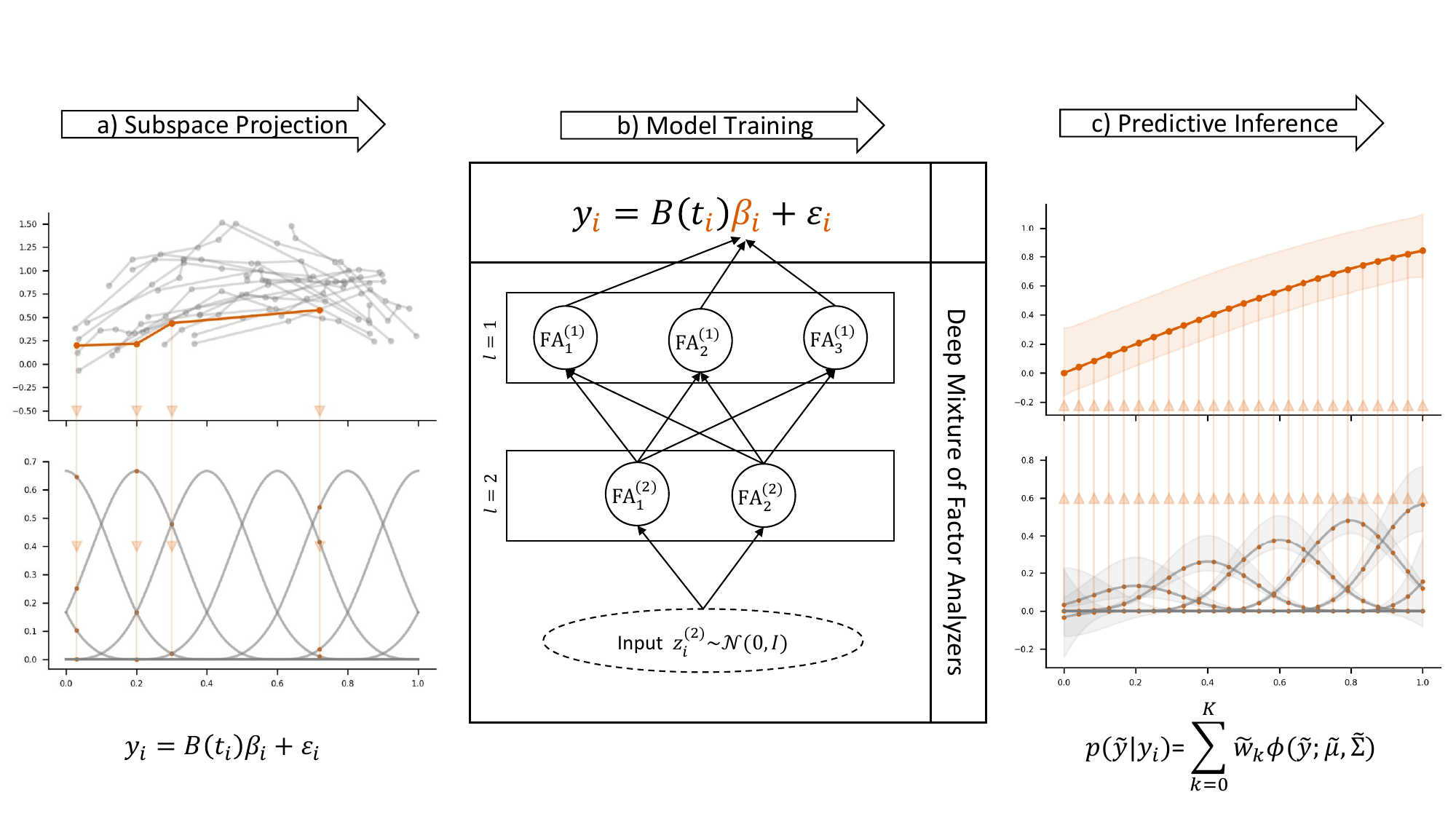}
\caption{ \small Schematic description of the DMLMM. a) represents the subspace projection. The number of observations $n_i$ as well as the time points $t_i$ may vary between individuals $i$. By means of a basis approximation, all vectors $y_i$ get projected into a lower dimensional sub-space of dimension $d$ controlled through the random coefficients $\beta_i$. b) represents the model training. The DMLMM consists of a regression layer of the form \eqref{eq: regression layer} and the DMFA prior for $\beta_i$. Here, we give an examplary  DMFA architecture with $L=2$ layers. The latent input variables $z_i^{(2)}$ are fed through a fully connected network with $L=2$ layers, with $K^{(1)}=3$ and $K^{(2)}=2$ components, respectively. The components of the network correspond to a factor analyzer  of form \eqref{eq: factormodel}. Each of the $K^{(1)}\cdot K^{(2)}=6$ possible paths through this model corresponds to a GMM component. c) represents posterior predicted inference based on the fitted DMLMM  for new unobserved data $\tilde{y}$ conditional on $y_i$ exploiting \eqref{eq: predictive distribution}.}
\label{fig: dmlmm}
\end{figure}

A key task that we address in these applications
is within subject prediction.  Suppose that for subject $i$ 
we need predictive inferences about unobserved data
$\tilde{y}$ at time points $\tilde{t}=\left(t_1,\dots,t_T\right)$. 
Integrating out $\beta$, the joint density of $(y_i,\tilde{y})$ 
given $\eta$ is a high-dimensional GMM,

\begin{align*}
    p(y_i,\tilde{y}|\eta)=\sum_{k=1}^K w_k \phi\left(\begin{bmatrix}y_i\\\tilde{y}\end{bmatrix};\begin{bmatrix}B(t_i)\mu_k\\B(\tilde{t})\mu_k\end{bmatrix},\begin{bmatrix}B(t_i)\Sigma_kB(t_i)^\top+\sigma^2I_{n_i} &B(t_i)\Sigma_kB(\tilde{t})^\top\\ B(\tilde{t})\Sigma_kB(t_i)^\top& B(\tilde{t})\Sigma_kB(\tilde{t})^\top+\sigma^2I_{T}\end{bmatrix}\right).
\end{align*}
leading to a conditional density for $\widetilde{y}$ given $y_i,\eta$ which is also a GMM:
\begin{align}\label{eq: predictive distribution}
    p(\tilde{y}\mid y_i,\eta)=\sum_{k=1}^K \tilde{w}_k \phi(\tilde{y};\tilde{\mu}_k,\tilde{\Sigma}_k),
\end{align}
where
\begin{align*}
    \tilde{w}_k&=\frac{w_k\phi\left(y_i;B(t_i)\mu_k,B(t_i)\Sigma_kB(t_i)^\top\right)}{\sum_{k=1}^Kw_k\phi\left(y_i;B(t_i)\mu_k,B(t_i)\Sigma_kB(t_i)^\top\right)}\\
    \tilde{\mu}_k &= B(\tilde{t})\mu_k-B(\tilde{t})\Sigma_kB(t_i)^\top\left(B(t_i)\Sigma_kB(t_i)^\top+\sigma^2I_{n_i} \right)^{-1}\left(y_i-B(t_i)\mu_k\right) \\
    \tilde{\Sigma}_k &= B(\tilde{t})\Sigma_kB(\tilde{t})^\top+\sigma^2I_{T}-B(\tilde{t})\Sigma_kB(t_i)^\top\left(B(t_i)\Sigma_kB(t_i)^\top+\sigma^2I_{n_i}\right)^{-1}B(t_i)\Sigma_kB(\tilde{t})^\top.
\end{align*}
Predictive inference can be obtained from \eqref{eq: predictive distribution} either in 
a plug-in fashion, using a point estimate of $\eta$, or by integrating
out the parameters over the posterior distribution or some 
approximation to it.  
In Section \ref{sec:post}, we will consider posterior
approximations and point estimates obtained using variational inference.
Figure~\ref{fig: dmlmm} illustrates the full DMLMM including the DMFA prior and model training process which we will discuss further next. 

\subsection{DMFA prior for subject specific random effects}\label{sec: dmfa}

The DMFA model was motivated by Viroli and McLachlan\citep{viroli2019deep} as a 
deep extension of the mixture of factor analyzers (MFA) model, which
can be thought of as a DMFA model with only one layer.  
While Viroli and McLachlan\citep{viroli2019deep} and Kock et al.\citep{KocKleNot2022} consider
DMFA models for multivariate data directly, here it will be
used as a prior for random effects in a LMM.  

The hierarchical DMFA prior is a generative model
for the random effects $\beta_i$ expressed in terms of 
latent variables arranged in a number of layers.  
Define $z_i^{(0)}:=\beta_i$, and 
write $\zil\in \mathbb{R}^{D^{(l)}}$, $i=1,\dots, n$, for 
latent variables at layer $l\in\{1,\dots, L\}$.
We define the model for $\zilm1$ $l=1,\dots, L$, 
in terms of $\zil$ as a mixture model, with $\Kl$ components.  
At level $l$, the mixing weights for the mixture 
are denoted $\wkl$, $k=1,\dots, K^{(l)}$, 
$\sum_k \wkl=1$.  The model for $\zilm1$ given 
$\zil$ is expressed generatively as follows:  for $l=1,\dots, L$, with 
probability $\wkl$, $\zilm1$ is generated as 
\begin{align}\label{eq: factormodel}
\zilm1 & = \mukl+\Bkl\zil+\epsikl,  
\end{align}
where $\epsikl\sim \ND(0,\delkl)$, $\mukl$ is a $\Dlm1$-vector, $\Bkl$ is a $\Dlm1\times \Dl$ lower triangular matrix, 
$\delkl=\diag(\delta_{k1}^{(l)},\dots, \delta_{k\Dlm1}^{(l)})$ is a $\Dlm1\times \Dlm1$ diagonal matrix with diagonal elements $\delta_{kj}^{(l)}>0$.  At the final layer
$z_i^{(L)}\sim \ND(0,I_{D^{(L)}})$. In the specification of the DFMA prior, we restrict the dimensionality of the latent variables to satisfy the Anderson-Rubin condition\citep{FruHosLop2024} $D^{(l+1)}\leq \frac{D^{(l)}-1}{2}$ for $l=0,\dots,L$, as it is a necessary condition for ensuring model identifiability. Figure~\ref{fig: dmlmm}b) gives an example for a DMFA prior architecture with $L=2$ layers. Kock et al.\citep{KocKleNot2022} recommend architectures with few layers and a rapid decrease in dimension. Following this recommendation, we consider models with $L=2$ layers throughout our experiments. 

Following the discussion of Viroli and McLachlan\citep{viroli2019deep}, the DMFA prior can be regarded as a GMM with $K=\prod_{l=1}^L K^{(l)}$ components.  The components correspond to ``paths'' through the factor mixture components at the different levels.  Write $k_l\in \{1,\dots, K^{(l)}\}$ for the index
of a factor mixture component at level $l$ and  let $k=(k_1,\dots, k_L)^\top$ index a path.  
Let $w_k=\prod_{l=1}^L w_{k_l}^{(l)}$,
$$\mu_k=\mu_{k_1}^{(1)}+\sum_{l=2}^L\left(\prod_{m=1}^{l-1}B_{k_m}^{(m)}\right)\mu_{k_l}^{(l)}
\;\;\mbox{ and }\;\;
\Sigma_k=\delta_{k_1}^{(1)}+\sum_{l=2}^L \left(\prod_{m=1}^{l-1} B_{k_m}^{(m)}\right) \delta_{k_l}^{(l)}\left(\prod_{m=1}^{l-1}B_{k_m}^{(m)}\right)^\top.$$
Then the DMFA prior corresponds to the Gaussian mixture density
$\sum_{k=1}^K w_k \phi(y;\mu_k,\Sigma_k).$

To get some intuition for the DMFA prior construction, 
it is helpful to consider the case of a single layer, $L=1$.  
In this case, the DMFA prior is a mixture of factor analyzers (MFA)
prior on the random effects.  Abusing notation by writing simply
$K=K^{(1)}$, $w_k=w_k^{(1)}$, $\mu_k=\mu_k^{(1)}$,
$B_k=B_k^{(1)}$, $\delta_k=\delta_k^{(1)}$, $k=1,\dots, K$, and $z_i=z_i^{(1)}$, $i=1,\dots, n$,
 \eqref{eq: factormodel}
specifies the prior for $\beta_i$ through the following 
single generative layer:  with
probability $w_k$, generate $\beta_i$ as
$$\beta_i=\mu_k+B_kz_i+\epsilon_{ik},$$
where $\epsilon_{ik}\sim \ND(0,\delta_k)$.  
Integrating out the latent variables $z_i$, the corresponding density
of $\beta_i$ is 
\begin{align*}
      \sum_{k=1}^K w_k \phi(\beta_i;\mu_k,B_kB_k^\top+\delta_k).
\end{align*} 
The low-dimensional latent variables $z_i$ allow a parsimonious
description of the dependence between the possibly high-dimensional
components in $\beta_i$;  conditionally on $z_i$, components of $\beta_i$
are independent.   
The latent variables $z_i$ are called factors, and the matrices $B_k$ are called factor loadings or factor loading matrices. The key idea of the DMFA prior is to replace the Gaussian assumption $z_i\sim \ND(0,I)$ with the assumption that the $z_i$'s themselves follow a MFA model.

In a Bayesian framework, Kock et al.\citep{KocKleNot2022} propose the following marginally independent priors for the parameters of a DMFA model, 
and we use similar priors for the hyperparameters on the DMFA prior
for the random effects. They use thick-tailed Cauchy priors on the component mean parameters $\mukl$ and half-Cauchy priors on the standard deviations $\delkl$. Thus integration over the model parameters yields a prior centered on zero for the random effect distribution, which does not introduce
an unwanted bias for $\beta_i$. In the DMLMM the same prior is used also for the standard deviation $\sigma$ of the error terms $\varepsilon_i$. For the component factor loading matrices $\Bkl$, they use the sparsity-inducing horseshoe prior of Carvalho and Polson\citep{CarPol2010}. Kock et al.\citep{KocKleNot2022} show that this prior choice is helpful with regularizing the estimation. Additionally, in the DMLMM  imposing sparsity on the factor loadings is motivated by the fact that the entries of the coefficient vector $\beta_i$ control local information and therefore each of the latent factors should control only a subset of components, but not the full vector. Typically the basis functions are chosen such that $B(t)$ is sparse as well. Lastly, the marginal prior for $w^{(l)}$ is a Dirichlet distribution allowing to select the number of clusters in a computationally thrifty
way, using overfitted mixtures \citep{RosMen2011}. A precise description of the priors is given in Web Appendix A.

\section{Posterior computation}\label{sec:post}
Next we review basic ideas of variational inference and explain how the scalable variational inference algorithm for the DMFA model in Kock et al.\citep{KocKleNot2022} can be extended to the new DMLMM with DMFA prior for the random effects.

\subsection{Variational inference}
Variational Inference (VI) \citep{BleKucMca2017}
learns an approximation to the posterior density
$p(\theta\mid y)$ in Bayesian inference using
an approximating family of densities 
$\{q_\lambda(\theta)$, $\lambda\in \Lambda\}$ 
where $\lambda$ are variational parameters to be chosen.  
The optimal approximation is obtained by finding 
the value $\lambda^*$ of $\lambda$ minimizing some measure
of dissimilarity between $p(\theta\mid y)$ and $q_\lambda(\theta)$.  
A common choice for the dissimilarity measure is
the reverse Kullback-Leibler (KL) divergence, 
$$\KL{q_\lambda(\theta)}{p(\theta\mid y))}=
\mathbb{E}_{q_\lambda}\left\lbrace\log(q_\lambda(\theta)/p(\theta\mid y)\right\rbrace,$$
where $\mathbb{E}_{q_\lambda}(\cdot)$ denotes expectation with
respect to $q_\lambda(\theta)$.  
Minimizing the reverse KL divergence is equivalent
to maximizing the Evidence Lower Bound (ELBO), 
\begin{equation}\label{eq: elbo}
\mathcal{L}(\lambda)=\mathbb{E}{q_\lambda}\left\lbrace\log(h(\theta))-\log(q_{\lambda}(\theta))\right\rbrace,
\end{equation}
where $h(\theta)=p(y\mid\theta)p(\theta)$. For the DMLMM
with a DMFA prior for the random effects, we consider variational
approximations leading to a closed form expression
for the ELBO.  
We optimize the ELBO using a stochastic gradient ascent (SGA) method 
which uses mini-batch sampling to effectively deal
with large datasets. We give a high level discussion of the approach next, a detailed discussion can be found in Kock et al.\citep{KocKleNot2022}.

\subsection{VI  for the DMFA} \label{sec: VI DMFA}
The SGA algorithm for the original DMFA model of Kock et al.\citep{KocKleNot2022} adapts stochastic VI \citep{HofBleWanPai2013} by partitioning  the variational parameters into ``global'' parameters $\lambda_{G}$, which parametrize variational posterior terms for shared model parameters such as the factor loading matrices $\Bkl$ or the component mean shift vectors $\mukl$, and ``local'' parameters $\lambda_{L}$, which 
parametrize variational posterior terms for observation-specific latent variables, such as $\zil$. Write $\lambda=(\lambda_G^\top,\lambda_L^\top)^\top$, and denote 
the value of $\lambda_L$ maximizing the ELBO for a 
given value of $\lambda_G$ as $M(\lambda_G)$.  
We then consider the ELBO as a function of $\lambda_G$, with
$\lambda_L$ fixed at $M(\lambda_G)$:
$$\overline{\cal L}(\lambda_G):=\mathcal{L}(\lambda_G,M(\lambda_G)).$$  
The stochastic VI algorithm we use optimizes $\overline{\cal L}(\lambda_G)$ 
where at step $m=1,\dots,M$ of the SGA algorithm there are two nested steps. First, the optimal local parameters $\widehat{\lambda}_{L}$ for the current global parameter vector $\lambda_{G}^{(m-1)}$ are updated. Then, the global parameters are updated as 
\begin{align} \label{eq sga update}
    \lambda_{G}^{(m)}=\lambda_{G}^{(m-1)}+a_m\circ \widehat{\nabla_{\lambda_G} \overline{\cal L}}(\lambda_{G}^{(m-1)}),
\end{align}
where $a_m$ is a vector-valued step size sequence, $\circ$ denotes elementwise multiplication, and 
$\widehat{\nabla_{\lambda_G} \overline{\cal L}}(\lambda_{G}^{(m-1)})$ is an unbiased estimate of the natural gradient 
\citep{amari98} of $\overline{\cal L}(\lambda_{G}^{(m-1)})$ based on a random data mini-batch, where $\overline{\cal L}(\cdot)$ denotes the ELBO with local parameters fixed at $\widehat{\lambda}_{L}$. Optimization of local variational parameters
is only required for the observations in the data mini-batch, 
which leads to an efficient algorithm for large data sets.  

\subsection{VI for DMLMMs}
The deep structure of the DMLMM  corresponds to a DMFA model with an additional regression layer of the form \eqref{eq: regression layer} on top (see Figure~\ref{fig: dmlmm}b)). The regression layer has a very similar structure to a single layer in the DMFA model, \eqref{eq: factormodel}, where the factor loading matrix is fixed at $B(t_i)$ and the mean shift vector is zero. This perspective allows us to extend the efficient VI scheme for DMFA to DMLMM as follows.

Let $\thetaDMFA$ denote the vector of all unknown model parameters for the DMFA prior and $\thetaReg=(\beta^\top,\sigma^2,\psi)^\top$ be the vector of the remaining parameters. The full set of unknown model parameters for the DMLMM  is then $\theta=(\thetaDMFA^\top,\thetaReg^\top)$. We assume a factorized variational approximation to the posterior density of the form 
\begin{equation}\label{eq: q}
q_\lambda(\theta)=q_\lambdaDMFA(\thetaDMFA)q_\lambdaReg(\thetaReg),  
\end{equation}
where $q_\lambdaDMFA(\thetaDMFA)$ is the density for
$\thetaDMFA$ with variational parameters $\lambdaDMFA$ and
\begin{equation*}
q_\lambdaReg(\thetaReg)=q(\sigma^2)q(\psi)\prod_{i=1}^nq(\beta_i),
\end{equation*}
where $q(\sigma^2)$ and $q(\psi)$ are inverse gamma densities and $q(\beta_i)$ is a multivariate Gaussian density with independent marginals. Then, 
\begin{align*}
    h(\theta)&=p(\theta)\prod_{i=1}^np(y_i\mid\theta)\\
    &=p(\thetaDMFA)p(\sigma^2\mid\psi)p(\psi)\prod_{i=1}^np(y_i\mid \beta_i,\sigma^2)p(\beta_i\mid\thetaDMFA),
\end{align*}
and \eqref{eq: elbo} can be decomposed as 
\begin{equation*}
    \mathcal{L}(\lambda)=\mathcal{L}^{\text{DMFA}}(\lambda)+\mathcal{L}^{\text{Reg}}(\lambda),
\end{equation*}
where 
\begin{equation*}
    \mathcal{L}^{\text{DMFA}}(\lambda)=\mathbb{E}{q_\lambda}\left[\sum_{i=1}^n\log(p(\beta_i\mid\thetaDMFA))+\log(p(\thetaDMFA))-\log(q_{\lambdaDMFA}(\thetaDMFA))\right]
\end{equation*} can be derived from the ELBO for the DFMA model
and 
\begin{equation*}
    \mathcal{L}^{\text{Reg}}(\lambda)=\mathbb{E}{q_\lambda}\left[\sum_{i=1}^n\log(p(y_i\mid\beta_i,\sigma^2))+\log(p(\sigma^2\mid\psi)p(\psi))-\log(q_\lambdaReg(\thetaReg))\right]
\end{equation*}
is available in closed form. More details on the calculation of ${\cal L}(\lambda)$  can be found in the Web Appendix B.

$\mathcal{L}(\lambda)$ has a similar structure to the ELBO for the DMFA model, where $\beta$ is an additional ``local'' parameter and $\sigma, \psi$ are ``global'' parameters. As a result, it is straightforward to adapt the updating approach explained in Section~\ref{sec: VI DMFA} to the DMLMM.

In the DMFA model the use of overfitted mixtures and ELBO values of short runs allows to choose a suitable architecture in a computationally thrifty way and this idea directly translates to the DMLMM.   The choice of the number of layers 
and factors in our DMLMM also follows the choices made in the DMFA model.  
Due to the parameter sharing, some components of the GMM representation \eqref{eq: gmm y} might be empty, even when there are data points assigned to every component in each layer \citep{SelGorJacBie2020}. While this does not affect the clustering induced by the DMFA prior it can have negative impact on the resulting density estimation. Hence, we recommend that after the full model is fitted the weights for empty components of the GMM density are manually set to zero and remaining weights are rescaled. 
Predictive inference in the DMLMM is carried out using the variational posterior mean as a point estimate for $\eta$.

\section{Real data illustrations}\label{sec:app}

In this section we showcase our DMLMM in diverse real data applications. First, we consider longitudinal CD4 counts, which are an established illustration in the longitudinal literature. Then, we consider a novel application on malaria transmission. Here, the deep structure of our approach is helpful to capture the complex temporal structure of the data. An application on missing data imputation for gene expression data is presented in Appendix~C.

\subsection{Longitudinal CD4 counts}

\subsubsection{Data and model description}
CD4 percentages are a popular prognostic marker of disease stage among human immunodeficiency virus (HIV)-infected individuals. Here, we consider data from the Multicenter AIDS Cohort Study \citep{KasOstDetPhaPolRin1987} which has been analyzed by many
previous authors \citep[e.g.~by][]{FanZha2000,WuChi2000,YaoMulWan2005}. The dataset contains repeated measurements for 283 MSM (men who have sex with men) who were tested HIV-positive between 1984 and 1991. Even though individuals were expected to get their measurements taken in regular $6$ month time intervals, the number of measurements and the measurement times differ per individual. The observed trajectories for all individuals are shown in Figure~\ref{fig: cd4_effects}a). 
 
The goal is to model the CD4 percentage trajectories in continuous time as well as to dynamically predict CD4 percentages at future time points. To this end, we denote by $y_i$ the $n_i$-dimensional vector of observed CD4 measurements on the probit scale for individual $i$. The design matrices $B(t_i)$ are constructed from $d=7$ Legendre  polynomials. Since $n_i$ varies greatly between individuals ranging from $1$ to $14$, $n_i<d$ for more than half of the individuals. Let $\tilde{y}$ denote the unobserved CD4 measurements on a fine equidistant grid $\tilde{t}$ over $[0,6]$ with $120$ grid points.

\subsubsection{Results} 
\begin{figure}[ht] 
\centering
\includegraphics[width=\textwidth,keepaspectratio]{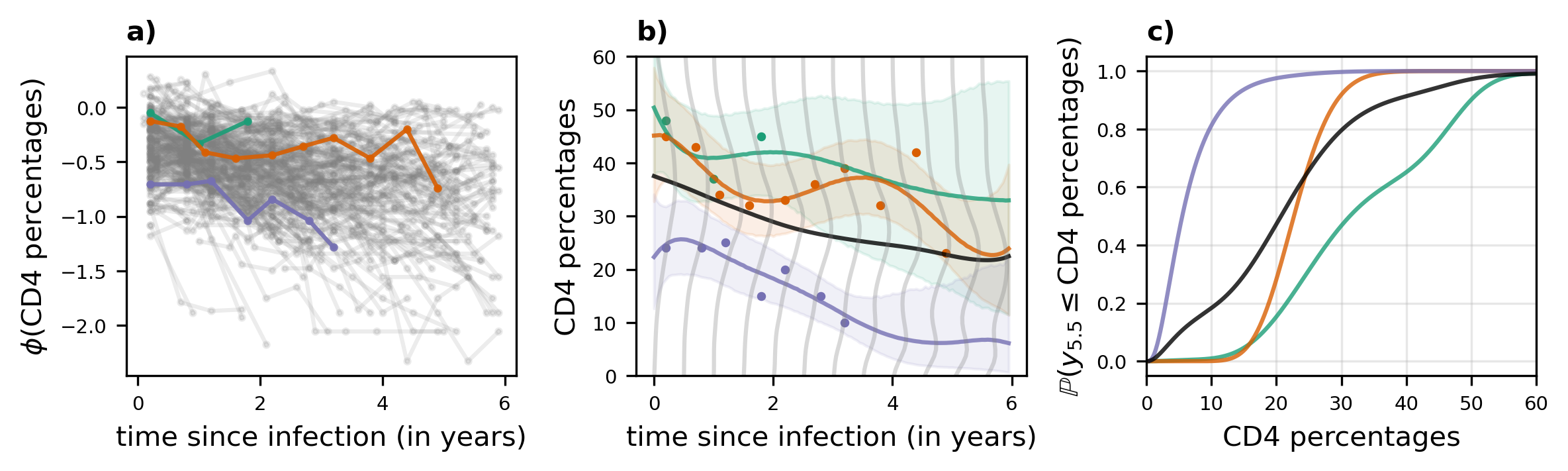} 
\caption{\small CD4 data. \textbf{a)} Spaghetti plot for all observed trajectories on the probit scale. Three randomly selected trajectories $y_i$ are marked in color. \textbf{b)} Predicted trajectories (bold) and $95\%$ credible intervals for the three randomly selected individuals, with observed measurements given by dots. The gray lines correspond to the estimated marginal densities $p(\tilde{y}\mid\hat{\eta})$ at different times $\tilde{t}$ and the mean  $\mathbf{E}(\tilde{y}\mid\hat{\eta})$ is given in black. \textbf{c)} Plot of the predicted CDF $\mathbb{P}(\tilde{y}_j\leq\cdot\mid y_i,\hat{\eta})$ at $\tilde{t}_j=5.5$ for the three individuals. The CDF of the marginal $\mathbb{P}(\tilde{y}_j\leq\cdot\mid\hat{\eta})$, $\tilde{t}_j=5.5$ is given in black.}
\label{fig: cd4_effects}
\end{figure}

Figure~\ref{fig: cd4_effects}b) shows the estimated mean effects $\mathbb{E}(\tilde{y}\mid y_i,\hat{\eta})$ with  $95\%$ pointwise credible intervals for three randomly selected individuals based on the observed measurements. Even in cases with limited measurement data, the method reconstructs meaningful trajectories by combining information from both the specific individual and the entire dataset. As expected, credible intervals are wider in regions where no measurements are observed and near the end of the time interval, where fewer data points are observed.
In a diagnostic context, within subject forecasting is of particular interest. By \eqref{eq: gmm y}, the cumulative distribution function (CDF) of a GMM can be expressed as a mixture of Gaussian CDFs. This allows for a simple calculation of the risk of the CD4 percentage of an individual falling below a threshold at a given time. Figure~\ref{fig: cd4_effects}c) shows the predicted CDFs $\mathbb{P}(\cdot \mid y_i,\hat{\eta})$ for three selected individuals at $\tilde{t}=4.5$. 

Further insight can be obtained through the predictive marginal density
for an individual for which no data has been observed, $p(\tilde{y}\mid\hat{\eta})$, 
which is depicted in Figure~\ref{fig: cd4_effects}b).  Computation
of this marginal density is simple as the mean effect, the variance function and the correlation function are available in closed  form
for a Gaussian mixture density. The mean effect shows an overall decreasing trend among individuals. The variance function is non-stationary and increases over time. The marginal distribution for time points near the end of the observation period becomes bimodal. As expected, time-points close to each other are estimated to be highly correlated. 

\subsection{Predicting malaria transmission in Afghanistan}\label{sec:sir}

The DMLMM approach is motivated by scenarios where both the number of observations and the dimension of the random effect is large. Such a scenario is commonly encountered when analyzing complex dynamical systems. Here, we reanalyze monthly data reporting malaria cases registered in Afghanistan from January 2005 to September 2015 \citep{AnwLewParPit2016}. In a recent analysis, Alahmadi et al.\citep{AlaFleCocDroKei2020} considered the data in a classical Bayesian parameter inference setting. Here we are interested in  forecasting  future case counts based on the observed data. 
There is an extensive literature on models for infectious diseases \citep{Het2000} with Susceptible–infected–recovered (SIR) models being a popular choice \citep{XuZhi2009,WuSteMoo2024}. 

White et al.\citep{WhiMauPonSarAguVanDayWhi2009}  and Alahmadi et al.\citep{AlaFleCocDroKei2020} propose a nonlinear ordinary differential equation (ODE) model based on the SIR model to describe the temporal population dynamics associated with malaria transmission. The model uses four coupled ODEs modelling four population compartments (uninfected and non-immune, infected with no prior immunity, uninfected with immunity and infected with prior immunity). These ODEs are highly parameterized to describe the complex evolution of the population compartments over time.  As none of the population compartments can be directly observed a fifth ODE describing the total number of treated cases is incorperated into the model. We observe $y_j\sim\ND(\log(c_j),\sigma^2)$, where $c_j$ denotes the number of new cases at time $t_j\in[0,T]$. 
A full description of the underlying latent ODE model can be found in Alahmadi et al.\citep{AlaFleCocDroKei2020}. We write $y_{t_1:t_2}$ for the vector of observations at time-points $t=(t_1,\ldots,t_2)^\top$ and $\theta$ for the vector of parameters of the ODE model.  

Since the ODE model is not fully observed, $p(y_{t:T}\mid \theta,y_{1:t})$ is not available in closed form and $p(y_{1:T}\mid\theta)$ is costly to evaluate as it involves numerically approximating a solution to the ODE. However, simulating data from the marginalized likelihood $p(y_{1:T})=\int p(y_{1:T}\mid\theta)d\theta$ is straight forward and the underlying model can be regarded as a black-box simulator.

While simulator-based or LFI methods such as Approximate Bayesian Computation (ABC)\citep{SisFanBea2018} are commonly used for parameter estimation and model comparison with computationally expensive likelihoods, predictive inference, such as computing the posterior predictive distribution of future observations or missing data, remains challenging due to the complexity of the underlying dynamics and the high dimensionality of the observations.
In contrast, the DMLMM enables closed form calculations of the predictive distribution without the need for direct inference on the model parameters, tedious calibration of hyperparameters, or selection of summary statistics based on expert knowledge. Instead,  the flexibility of the DMLMM allows us to learn a low-dimensional representation of the high-dimensional observations that captures the relevant information for prediction.

\subsubsection{Experimental design} As the uninformative prior $p(\theta)$ used in Alahmadi et al.\citep{AlaFleCocDroKei2020} results in many unrealistic time series, we reject any simulated time series for which the number of simulated cases never exceeds $100$. Since we regard $p(y)=p(y_{1:128})$ as a black box simulator, we do not need  to make this constraint explicit in the model formulation. We generate $7,500$ samples from $p(y)$, which we split into a training set with $5,000$ samples and a test set with $2,500$ samples. Our goal is to approximate  $p(y_{81:128}\mid y_{1:80})$ using  the joint samples from $p(y_{1:128})$. Järvenpää and Corander\citep{JarCor2023} discuss how ordinary ABC can be used in this setting and we use their approach, which we label ABC, as a benchmark. The design matrices $B(\cdot)$ used for DMLMM incorporate a $20$-dimensional spline basis, with $6$ splines modelling a yearly seasonality to account for the seasonal forcing associated with malaria transmission, and the remaining basis functions modelling an additive trend. 

\subsubsection{Results} 
\begin{figure}[ht] 
\centering
\includegraphics[width=\textwidth,keepaspectratio]{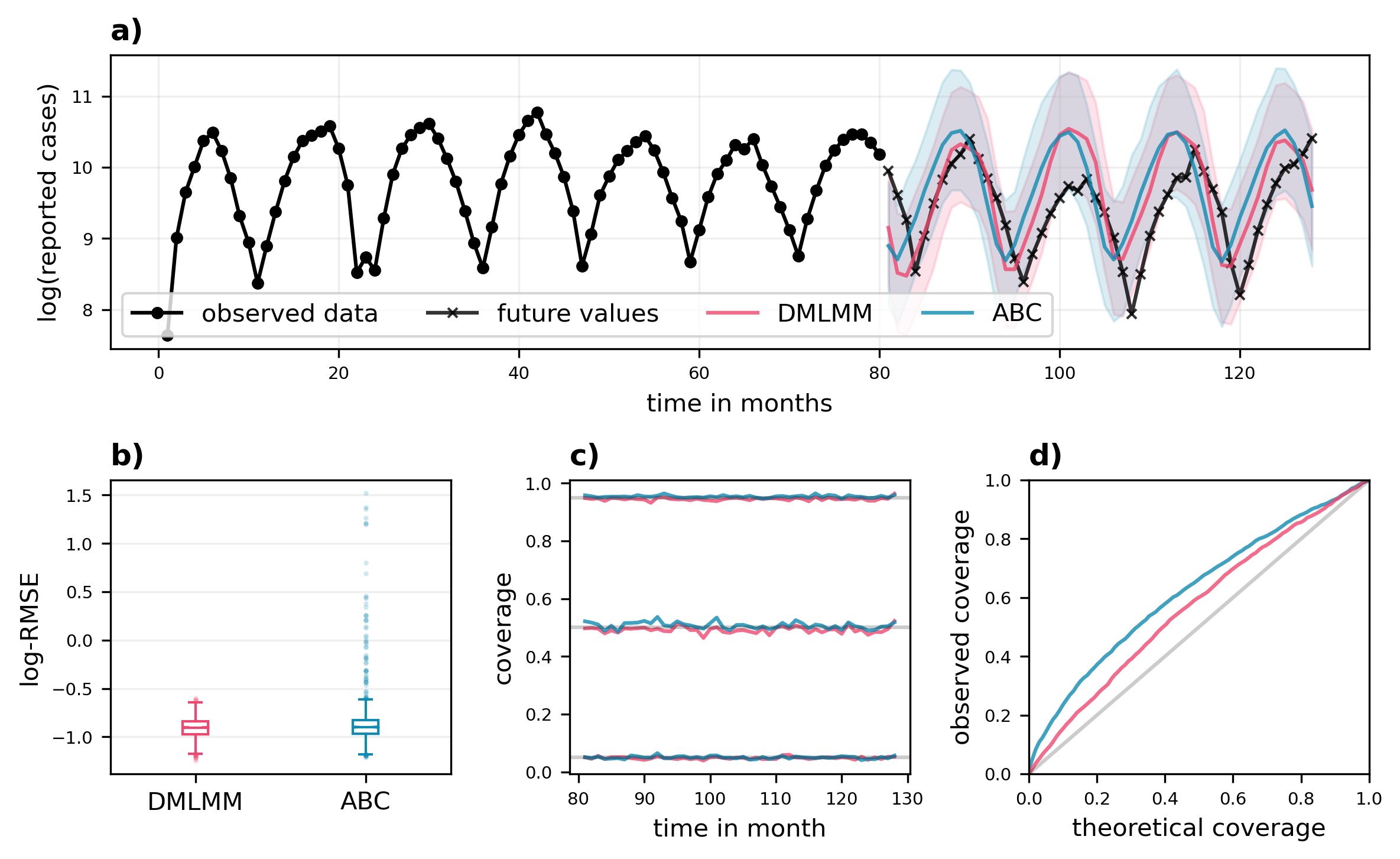} 
\caption{\small Malaria data. \textbf{a)} Prediction of the real, observed time series on registered malaria cases in Afghanistan. Shown is the predictive mean (bold) as well as a 95\% credible interval for DMFA (red) and ABC (blue). \textbf{b)} Boxplots for the logarithmic RMSE   across 2,500 independent realizations from the true model for DMFA (left) and ABC (right). \textbf{c)} Observed coverage rates for pointwise $95\%$, $50\%$ and $5\%$ credible intervals for DMFA (red) and ABC (blue). \textbf{d)} Observed coverage of elliptical credible sets from the $48$-dimensional posterior predictive distribution $p(y_{t+1:T}\mid y_{1:t})$.}
\label{fig: sir}
\end{figure}

Figure~\ref{fig: sir}a) shows the predicted time series for the observed data. Both, ABC and the DMLMM recover the general behaviour of the unobserved data points well. Studying the 95\% credible intervals for both methods shows no large difference between
the DMLMM approach and ABC, although our approach has slightly
better coverage properties.

The DMLMM also performs slightly better in terms of the root mean square error (RMSE) $\sqrt{\frac{1}{T-t}\sum_{t'=t}^T(y_{t'}-\widehat{y}_{t'})^2}$ with a mean of $0.41$ (ABC: $0.43$), median of $0.4$ (ABC: 0.41) and a standard deviation of $0.04$ (ABC: $0.2$) across all repetitions from the test data, as summarized  in Figure~\ref{fig: sir}b). Here, $\widehat{y}_{t'}$ are the posterior predictive mean estimates for ${y}_{t'}$. While the pointwise credible intervals for both methods seem very well calibrated at levels $0.05,0.5,0.95$ (Figure~\ref{fig: sir}c)), observed coverage rates of elliptical credible sets from the $48$-dimensional predictive distribution $p(y_{81:128}\mid y_{1:80})$  are closer to  the nominal levels for the DMLMM as shown in Figure~\ref{fig: sir}d). 

\subsubsection{Prior-data conflict} Recently, Nott et al.\citep{NotWanEvaEng2020} proposed a method for detecting prior-data conflicts in Bayesian models based on comparing prior-to-posterior Rényi divergences of the observed data with the prior-to-posterior divergence under the prior predictive distribution for the data. Since the marginal distribution $p(y_{t+1:T})$ acts as a prior to the implicit likelihood $p(y_{1:t}\mid y_{t+1:T})$, these checks translate directly to the predictive model described above. A tail probability for a model check can be computed as $p=\mathbb{P}\left[ G(y_{1:t})\geq G(y^{(\text{obs})}_{1:t})\right]$, where $G(y_{1:t})=\KL{p(y_{t+1:T}\mid y_{1:t})}{p(y_{t+1:T})}$. A small $p$-value indicates that the observed data is surprising under the assumed model and  Chakraborty et al.\citep{ChaNotEva2022} discuss how $p$ can be estimated in likelihood-free models through GMM-approximations. It is straightforward to translate their approach to the DMLMM, as the  DMFA prior allows for closed form GMM approximations of all quantities necessary to calculate $p$. The tail probability is estimated as $p=0.0024$ indicating that the latent ODE-model might need to be reexamined. This result is in line with the posterior-predictive checks for the malaria data conducted in  Alahmadi et al.\citep{AlaFleCocDroKei2020}.

\section{Simulation}\label{sec:sim}
To further illustrate in which scenarios the deep structure of the DMLMM is beneficial we consider three distinct simulation set-ups motivated by real world applications. We compare our DMLMM method to several established benchmarks, including a (non-deep) mixture of linear mixed models fitted by expectation maximization (MLMM), a random coefficient model (LMM), a mixture of linear models (MLM) and functional principal component analysis (FPCA).

\subsection{Simulation design} 
We consider the following three different data generating processes (DGPs). For each DGP we draw $50$ independent datasets.

\textbf{DGP 1}: We reanalyze the simulation study conducted in Wang et al.\citep{WanFarGupZhen2021}. In particular, for $i=1,\dots,600$, draw $0\leq t_1,\dots,<t_{10}\leq1$ uniformly on $[0,1]$, $g_i\sim \UD\{-1,1\}$ and $\xi_{i1}\sim\ND(0,0.1^2)$, $\xi_{i2}\sim\ND(0,0.045^2)$, $\xi_{i3}\sim\ND(0,0.01^2)$, $\xi_{i4}\sim\ND(0,0.001^2)$. Then, 
$$y_{ij}=g_i\sin(4\pi t_{ij})+\sqrt{2}\sum_{k=1}^4\xi_{ik}\sin(k\pi t_{ij})+\varepsilon_{ij},$$
where $\varepsilon_{ij}\sim\ND(0,0.3^2)$, $j=1,\dots,10$. 
This data set contains only two groups with means $\pm \sin(4\pi t)$ and observation specific functional errors $\sqrt{2}\sum_{k=1}^4\xi_{ik}\sin(k\pi t)$. Observation specific errors based on a truncated Karhunen-Loève expansion are often considered in the analysis of biomedical functional data \citep{MarInaKin2021}. 

\textbf{DGP 2}:
We consider $i=1,\dots,100$ observations of the form $y_{ij}=f_i(t_{ij})$ for $n_i\sim\UD\{15,16,\dots,25\}$ random timepoints $t_{ij}\in[10,20]$, where $f_i(t)$ is a solution to the following system of stochastic differential equations describing a Van der Pol oscillator
\begin{align*}
    \frac{d}{dt}f(t)&=g(t)+0.5\frac{d}{dt}W_{if}(t)\\
    \frac{d}{dt}g(t)&=\theta_i(1-f(t)^2)g(t)-f(t)+0.5\frac{d}{dt}W_{ig}(t),
\end{align*}
with $f(0)=1$, $g(0)=0.1$. $W_{ig}$ and $W_{if}$ are independent Brownian motions incorporating complex randomness into the observations. This enforces a complex dependence structure between nearby time points, for which the 
DMLMM is misspecified. Additionally, the parameter $\log(\theta_i)\sim\UD(1,5)$ has a continuous prior so that the observations cannot be easily separated into distinct groups. DGP 2 has a similar structure to the malaria model analyzed in Section~\ref{sec:sir}. 

\textbf{DGP 3}: This DGP is motivated by missing value imputation in time-course gene expression studies and related to experiments conducted by  Mao and Nott\citep{MaoNot2021}. Let 
$$y_{ij}=\beta_{i1}\cos\left(w_{i1}\pi \frac{t-1}{39}\right)+\beta_{i2}\sin\left(w_{i2}\pi \frac{t-1}{39}\right)+\varepsilon_{ij},\qquad i=1,\dots,120; j=1,\dots,40$$ where $\varepsilon_{ij}\sim\ND(0,0.1^2)$ is iid noise, and $\beta_{i1},\beta_{i2}\sim\UD\{1,0.1\}$, $w_{i1}\sim\UD\{1,2,3\}$, $w_{i2}\sim\UD\{7,8,9\}$ are parameters controlling the temporal trend. In each row of the matrix $(y_{ij})_{ij}$ 15 to 20 randomly selected data
points are removed. This data set contains $36$ clusters, some of which are difficult to distinguish, and a comparable small number of observations. 

\begin{figure}[ht] 
\centering
\includegraphics[width=\textwidth,keepaspectratio]{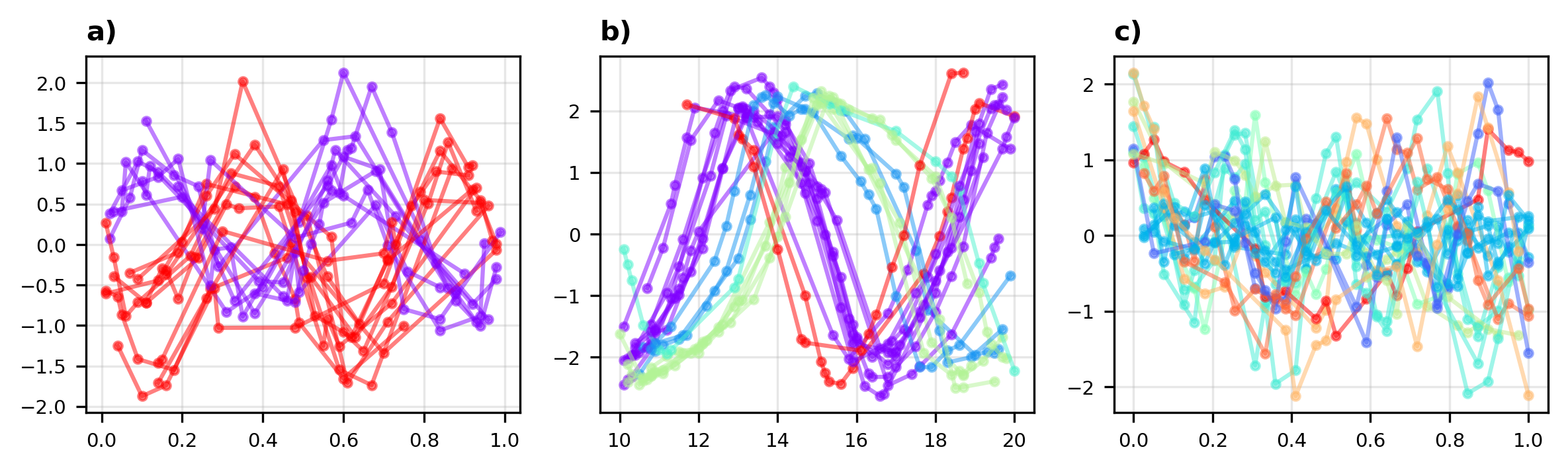} 
\caption{\small Simulations. 20 draws from DGP 1 -- DGP 3 (\textbf{a)}--\textbf{c)}). The colors correspond to the implicit clustering by DMLMM. Trajectories from the same cluster have the same color.}
\label{fig:sim_clusters}
\end{figure}

\subsection{Results} We consider $d=10$ basis functions for each DGP. Here, $1,000$ iterations of the SGA algorithm for DMLMM take about $4$ minutes on a standard laptop. Figure~\ref{fig:sim_clusters} shows simulations from the three DGPs. The GMM structure of the DMLMM approximation facilitates an implicit clustering of the subjects $y_i$ and the clustering for one run is highlighted by color. Notably, DMLMM recovers the two clusters for DGP 1 well. For both DGP 2 and DGP 3 a large number of Gaussian components is utilized. There is no ground truth clustering for DGP 2 available, but DMLMM groups trajectories with similar shapes in a meaningful way. 
Performance is evaluated in terms of the RMSE for the predictive distributions $p(\tilde{y}\mid y_i)$ and the negative log-score $-\log(p(y_i))$ evaluated on an additional hold-out test set. The summarized results in Table~\ref{tab:simulation} indicate that DMLMM performs robustly across all datasets and is competitive when compared to benchmark methods. On DGP 1 DMLMM is slightly outperformed by the non-deep MLMM. DGP 1 has only two Gaussian components, so the deep structure of DMLMM might not be fully leveraged. Conversely, on the complexer data sets DGP 2 and DGP 3 DMLMM is the best performing method. Both, DGP 2 and DGP 3 exhibit intricate temporal trends necessitating a complex random effects distribution. This is precisely the scenario DMLMM is tailored for. Moreover, DMLMM has superior performance in terms of density estimation for unobserved data, as measured by the log-score. This indicates that the DMLMM predictive distributions are useful for capturing predictive uncertainty, which is important for both  prediction as well as other purposes such as the prior-data conflict checks considered in Section~\ref{sec:sir}.

\begin{table}[ht]
\resizebox{\textwidth}{!}{\begin{tabular}{c|cc|cc|cc}
 & \multicolumn{2}{c|}{DGP 1} & \multicolumn{2}{c|}{DGP 2} & \multicolumn{2}{c}{DGP 3} \\
 & log-RMSE & neg. log-score & log-RMSE & neg. log-score & log-RMSE & neg. log-score \\ \cline{2-7} 
DMLMM & -1.54 (0.34) & 7.09 (0.27) & \textbf{-1.66 (0.45)} & \textbf{6.49 (1.40)} & \textbf{-1.08 (0.19)} & \textbf{14.93 (0.95)} \\
MLMM & \textbf{-1.62 (0.27)} & \textbf{6.33 (0.18)} & -1.55 (0.44) & 9.06 (2.24) & -1.06 (0.19) & 16.06 (0.99) \\
LMM & -1.18 (0.37) & 8.85 (0.06) & -1.52 (0.50) & 9.19 (0.53) & -0.94 (0.23) & 20.43 (0.51) \\
MLM & -1.43 (0.50) & 8.14 (0.23) & -0.73 (0.46) & 17.99 (1.27) & -0.84 (0.32) & 18.85 (0.78) \\
FPCA & -1.02 (0.24) & 8.97 (0.10) & -0.81 (0.30) & 16.76 (0.89) & -0.70 (0.37) & 21.61 (0.60)
\end{tabular}}
\caption{\small Simulations. Log-RMSE and negative log-score for the three DGPs (columns) for the five benchmark methods considered. The mean and the standard deviation (in brackets) are reported rounded to two digits. The best value across each column is marked.}
\label{tab:simulation}
\end{table}

\section{Conclusion and discussion}\label{sec:disc}

In this paper, we have introduced the DMLMM, which leverages the DMFA model as a prior for the random effects distribution. Our approach complements existing literature on models for complex longitudinal data and it is particularly suited for high dimensional settings. We demonstrate the effectiveness of the approach in simulations and biomedical applications in various scenarios, including within-subject prediction for unbalanced longitudinal data, LFI, and missing data imputation. Our DMLMM outperforms existing methods in these applications. While our focus has been on longitudinal data analysis, the DMLMM framework can be applied in other domains, including functional data analysis and Bayesian nonparametrics and it is a flexible model for researchers across different fields. While we have focused on temporal trends, many applications involve covariates that can influence the response. Extending the DMLMM to accommodate covariate-dependent effects is a further direction for future research.

\FloatBarrier
\bibliography{bib}

\end{document}